\documentclass[conference]{IEEEtran}
\IEEEoverridecommandlockouts
\usepackage{cite}

\usepackage{comment}
\usepackage{url}
\usepackage{multirow}
\usepackage{makecell}
\usepackage{booktabs}
\usepackage{amsmath,amssymb,amsfonts}
\usepackage{algorithmic}
\usepackage{graphicx}
\usepackage{textcomp}
\usepackage{xcolor}
\usepackage{threeparttable}
\usepackage[strings]{underscore}
\usepackage{amsmath,amssymb,amsfonts}
\usepackage{algorithmic}
\usepackage{graphicx}
\usepackage{textcomp}
\usepackage{xcolor}
\def\BibTeX{{\rm B\kern-.05em{\sc i\kern-.025em b}\kern-.08em
    T\kern-.1667em\lower.7ex\hbox{E}\kern-.125emX}}
\begin{document}

\title{DAM: A Universal Dual Attention Mechanism for Multimodal Timeseries Cryptocurrency Trend Forecasting

\thanks{Luyao Zhang is supported by the National Science Foundation China (NSFC) on the project entitled "Trust Mechanism Design on Blockchain: An Interdisciplinary Approach of Game Theory, Reinforcement Learning, and Human-AI Interactions (Grant No. 12201266). Yihang Fu is supported by the 2023 Student Experimental Learning Fellowship (SELF), Duke Kunshan University.}
}

\author{
  \IEEEauthorblockN{%
    Yihang Fu\IEEEauthorrefmark{2}
    Mingyu Zhou\IEEEauthorrefmark{2}
    Luyao Zhang\IEEEauthorrefmark{2}\textsuperscript{*}\thanks{\textsuperscript{*} The corresponding author: Luyao Zhang (lz183@duke.edu), Data Science Research Center and Social Science Division, Duke Kunshan University.}
  }
  \IEEEauthorblockA{%
    \IEEEauthorrefmark{2}Duke Kunshan University, 8 Duke Ave., Suzhou, China, 215316
  }
}

\maketitle

\begin{abstract}
In the distributed systems landscape, Blockchain has catalyzed the rise of cryptocurrencies, merging enhanced security and decentralization with significant investment opportunities. Despite their potential, current research on cryptocurrency trend forecasting often falls short by simplistically merging sentiment data without fully considering the nuanced interplay between financial market dynamics and external sentiment influences. This paper presents a novel Dual Attention Mechanism (DAM) for forecasting cryptocurrency trends using multimodal time-series data. Our approach, which integrates critical cryptocurrency metrics with sentiment data from news and social media analyzed through CryptoBERT, addresses the inherent volatility and prediction challenges in cryptocurrency markets. By combining elements of distributed systems, natural language processing, and financial forecasting, our method outperforms conventional models like LSTM and Transformer by up to 20\% in prediction accuracy. This advancement deepens the understanding of distributed systems and has practical implications in financial markets, benefiting stakeholders in cryptocurrency and blockchain technologies. Moreover, our enhanced forecasting approach can significantly support decentralized science (DeSci) by facilitating strategic planning and the efficient adoption of blockchain technologies, improving operational efficiency and financial risk management in the rapidly evolving digital asset domain, thus ensuring optimal resource allocation.

\end{abstract}


\section{Introduction}
As an advanced product of distributed systems, Blockchain has shown its potential in various application scenarios for its enhanced security, transparency, and decentralization. Cryptocurrency, as a decentralized digital currency empowered by Blockchain, inherits the features of Blockchain and shows great investment value. The increasing market volume and value of cryptocurrencies have attracted significant attention from investors. However, the high level of volatility associated with these assets not only raises concerns about risk but also presents challenges for investors trying to make forecasts. In this case, researchers tried statistical and machine learning methods to predict the volatility (see Table~\ref{Related Work}). Recently, with the support of language models, researchers found the effect of public sentiment on the financial market~\cite{li2020multimodal}. Cutting-edge deep learning methodologies have emerged as formidable tools for multimodal stock market predictions~\cite{bhatt2023machine,sardelich2018multimodalstock,shin2019deep,zhang2018improving}. This perspective naturally extends to cryptocurrency market forecasting and intuitively will lead to a better forecasting performance~\cite{anamika2023does,sapkota2022news}.

However, most of the current research only simply concatenates the sentiment data but ignores the intrinsic influence between the financial market data and the external sentiment data~\cite{raju2020real,parekh2022dl}. Also, researchers often neglected either news sentiment or social media sentiment which represent respectively objective and subjective sentiment data~\cite{abraham2018cryptocurrency,kraaijeveld2020predictive,rognone2020news,huy2019predicting}. While the current consensus acknowledges the boost sentiment provides to predictive capabilities, neglecting a comprehensive data fusion could introduce prediction biases, given that investors often overreact to certain offline events, as highlighted in previous studies~\cite{fu2022ai,antweiler2006us}. 

To make the Blockchain and cryptocurrency community more engaged, it is necessary to propose an explainable and efficient method that could be used in more financial forecasting scenarios. In this case, we are aiming to fusion the sentiment data naturally and to investigate if it could work with other financial indicators to enhance the forecasting performance. We sourced Bitcoin data via the Cryptocompare API, extracted Bitcoin news sentiment from Nasdaq, and social media data from a public Kaggle dataset. We designed DAM, a universal Dual Attention Mechanism for multimodal financial time-series data fusion. This mechanism managed to capture the intramodal and the crossmodal information and was tested to be efficient in enhancing the forecasting ability of our pre-trained LSTM model. We also did an Ablation Study to make our data fusion mechanism more explainable. At last, we compared our model with other popular models used for forecasting the cryptocurrency market. Key contributions of our research can be concluded as follows:
\begin{enumerate}
    \item Our Multimodal fusion does enhance the efficacy of deep learning algorithms. Our comparative study shows that our dual attention mechanism increases a 20 percent improvement.
    \item Our Ablation study shows that our dual attention mechanism successfully integrates both intramodal information and cross-modal information.
    \item Our research is a novel interdisciplinary approach bridging distributed systems, Natural Language processing, and financial trend and volatility forecasting. This research is expected to make the distributed system-related product more accessible and explainable.
\end{enumerate}

The rest of the paper is organized as follows:
\begin{itemize}
\item Section~\ref{Related Work} introduces the foundational concepts necessary for understanding Blockchain, Cryptocurrency, sentiment analysis, financial market forecasting, and multimodal temporal data fusion techniques.
\item Section~\ref{Methodology} describes our data sources, data preprocessing methods, and the architecture of our Dual Attention Mechanism (DAM) model.
\item Section~\ref{Results} presents our experimental results, including comparative studies and ablation analyses.
\item Section~\ref{discussion} discusses the enhanced performance and limitations of our model, with further potential improvements and future research directions outlined in Section~\ref{conclusion and future work}.
\end{itemize}
In the appendix~\ref{appendix}, we provide more detailed information about the historical-comparative study and some benchmark models we used in our experiment.

\textbf{Data and Code Availability Statement}: Our data is released under open access terms on Harvard Dataverse~\cite{DVN/HWBT53_2024}, and our code is available on GitHub at \url{https://github.com/KerwinFuyihang/Multimodal_crypto_prediction}.

\section{Background}
\label{Related Work}
This section introduces the background knowledge of our article. It begins with an introduction to Blockchain as an advanced distributed system and the great impact of cryptocurrency in the financial market. We then explore how sentiment influences cryptocurrency trends and review the predominant sentiment analysis methods. Finally, we summarize financial forecasting models, focusing on multimodal fusion methods that incorporate sentiment data into forecasting tasks.
\subsection{Distributed System, DeSci, Blockchain, and Cryptocurrency}
A distributed system is an interconnected network of independent computers that work together to perform a unified task. This architecture, characterized by its decentralized nature, ensures greater fault tolerance, scalability, and resource sharing. By leveraging multiple computers, a distributed system distributes computation and data storage, enhancing efficiency and reliability.

Decentralized science (DeSci) builds upon these principles of distributed systems. It applies the decentralized, collaborative framework to scientific research, promoting open access, transparency, and collective contribution. This integration highlights how advanced computing methodologies are revolutionizing traditional scientific processes.

Blockchain is a distributed system (also an application of DeSci) with properties of enhanced security, transparency, and decentralization, where many of its fundamental algorithms are adopted from classic distributed systems~\cite{herlihy2019blockchains}. Unlike traditional databases, blockchain stores data across a network of computers, making it highly resistant to unauthorized changes and hacks. Each block in the chain contains a number of transactions, and every time a new transaction is added, a record of that transaction is added to every participant's ledger~\cite{hashemi2020cryptocurrency}. This decentralized nature ensures no single point of failure, enhancing security and resilience. Moreover, the transparency of blockchain allows every participant to view the entire history of transactions, ensuring data integrity and trust among users.

Therefore, Blockchain-based technologies have been proposed in different scenarios, including data management, data privacy, transaction efficiency,etc~\cite{hu2023data, shae2018transform, ng2021ldsp}. Cryptocurrency is the product empowered by Blockchain shown to have the above features and shows potential as new transaction payments. Unlike traditional currencies, cryptocurrencies are decentralized. This decentralization reduces the risk of systemic failures and central control, making the system more resilient and less prone to censorship or manipulation. Furthermore, cryptocurrencies enable rapid cross-border transactions with lower fees compared to traditional banking systems. This is particularly beneficial for international remittances and trade. Also, the cryptocurrency's vast market capitalization indicates substantial investor confidence and interest in these digital assets. Prominent cryptocurrencies, notably Bitcoin and Ethereum, have demonstrated exceptional investment value. The market value of these and other popular cryptocurrencies has experienced a huge surge, increasing by thousands of times. 

However, the great volatility of the cryptocurrency market also raises concerns about investment risks. In December 2017, Bitcoin's value increased by approximately 2,700 percent, reaching an unprecedented high of 19,891 USD. In the same year, other cryptocurrencies even surpassed Bitcoin in terms of growth rate. However, beginning in January 2018, a widespread sell-off of cryptocurrencies ensued, leading to a steep 65 percent drop in Bitcoin's price within just a month. This extreme volatility has captured the attention of researchers, who are now increasingly focused on developing better methods to forecast market trends throughout the volatility.
\subsection{Sentiment Analysis on Cryptocurrency Topics}

Researchers respectively examined the impact of news sentiment and people's reactions on social media to the stock and cryptocurrency markets~\cite{chen2023poster,almeida2023systematic}. They demonstrated individuals' susceptibility to negative sentiments, thereby impacting cryptocurrency prices. Consequently, positing that integrating news and social media sentiments could enhance cryptocurrency price prediction seems rational. Scholars predominantly employed the VADER (Valence Aware Dictionary and sEntiment Reasoner) method, a widely adopted lexicon-based sentiment analysis tool for sentiment analysis in cryptocurrency research~\cite{hutto2014vader, gurrib2022predicting, critien2022bitcoin}. However, VADER's lexicon is primarily designed for general sentiment analysis and may lack the specificity required to capture the nuances and idiosyncrasies of crypto-related texts. The lexicon may not include the evolving and dynamic terminology, potentially leading to inaccurate sentiment analysis. More language models have already been introduced to do the sentiment analysis and a pre-trained model on a specific topic was tested to show especially outstanding performance~\cite{min2023recent}. In this article, we introduce a pre-trained BERT model based on over 3 million crypto-related texts called CryptoBERT released online, and scholars have shown success in applying this model to their research~\cite{passalis2022multisource}.

\subsection{Financial Time Series Forecasting}
This section reviews prior research in time series forecasting within the domains of stock and cryptocurrency markets. The main reviews are listed in Table~\ref{review}. It is important to note that the presented selection is limited to exemplary research papers showcasing cutting-edge models. We found that LSTM-based architectures are the most welcomed and accurate among all the forecasting models in this area. Even though transformers have shown outstanding performances in literature~\cite{sridhar2021multi,muhammad2023transformer,yoo2021accurate} as well, researchers found that the transformers-based model failed to tackle temporal information all the time~\cite{zeng2023transformers}. More discussions of the underperforming transformers-based models will be presented in section~\ref{discussion}. It is worth knowing that outcome discrepancies among similar model types can be attributed to variations in model architectures, dataset characteristics, and fusion methodologies.

\subsection{Temporal Data Fusion}
In this article, we regard the classic financial data and the sentiment data as separate modalities. A typical modal fusion method is simple concatenation~\cite{shahi2020stock,boukhers2022ensemble} $A^{N \times T} \bigoplus B^{M \times T}$, where T is the number of time steps, N and M are the number of features in certain modalities. However, simple concatenations often fail to consider the inherent inter-modal information~\cite{stahlschmidt2022multimodal}. Another fundamental method for combining modalities involves either additive or multiplicative operations. These processes are versatile, enabling the integration of dual data streams into almost any singular machine learning framework. For unimodal data or features, denoted as $x_1$ and $x_2$, the additive approach synthesizes a new merged representation $z_{mm}$, expressed as $z_{mm} = w_0 + w_1x_1 + w_2x_2 + \epsilon$. In this equation, $w_1$ and $w_2$ are the learned weights for merging $x_1$ and $x_2$, with $w_0$ as the bias, and $\epsilon$ representing the error. Viewing $z_{mm}$ as the preliminary prediction $\hat{y}$, this additive fusion mirrors the concept of late or ensemble fusion, exemplified by $\hat{y} = f_1(x_1) + f_2(x_2)$, where $f_1$ and $f_2$ are individual unimodal predictors. The multiplicative approach, considered a second-order extension to the additive method, incorporates an interaction term $w_3(x_1 \times x_2)$~\cite{liang2022foundations}. Thus, the combined model is given by $z_{mm} = w_0 + w_1x_1 + w_2x_2 + w_3(x_1 \times x_2)$.

To effectively understand input-dependent representational information, experts have turned to attention-based strategies~\cite{fu2021cross,zhou2021information,shih2019temporal}. Such methods are typically encapsulated as $z_{mm} = x_1 \odot h(x_2)$, with $h$ signifying a function that uses sigmoid activation, and $\odot$ indicating an element-wise multiplication. The function $h(x_2)$, often termed 'attention weights', is derived from $x_2$ and applied to selectively emphasize aspects of $x_1$. In neural network architectures, particularly in Transformer models, the Query-Key-Value (QKV) attention mechanism is essential~\cite{lim2021temporal,zhang2022temporal}. It calculates attention weights based on similarity measures between queries ($Q$) and keys ($K$). The attention weights ($A$) are computed using a softmax function on the dot products of $Q$ and $K$.

Furthermore, both Hidden Markov Models (HMM) and Bayesian Networks have been employed as data modeling methods~\cite{hashish2019hybrid}. Contemporary research has also embraced the application of Neural Networks for data fusion~\cite{costa2023show,zhang2022forecasting}. In contrast to early data fusion strategies, recent methodologies have introduced late fusion (late model fusion) techniques to integrate multiple machine learning models~\cite{bhatt2023machine,critien2022bitcoin}. Moreover, researchers proposed an alternative approach, utilizing Neural Tensor Network for event embedding learning, a departure from conventional text-based sentiment analysis~\cite{ding2015deep}. This method replaced the traditional idea of doing text-based sentiment analysis. Also, researchers found that the stock movement cannot be assumed to be influenced by one-to-one events~\cite{li2020multimodal}. That is, news or social media information on other stocks or cryptocurrencies could cause the co-movements of the cryptocurrency.

\section{Methodology}
\label{Methodology}
This section introduces the methodology part of this article. We first introduce our data sources and the CryptoBERT model for the sentiment data preprocessing. Then, we explain and showcase the structures of our Dual Attention Mechanism (DAM) for Multimodal Data Fusion in financial forecasting. At last, we introduce the evaluation metrics used in our experiments.
\subsection{Data Acquisition}
We queried cryptocurrency data (Bitcoin) ranges on a daily level from 2020-01-01 to 2023-06-28 by using CryptoCompare API\footnote{https://min-api.cryptocompare.com/}.

Our paper plans to incorporate social media information and news information to do the multivariate time series cryptocurrency forecasting. The social media posts are collected from Kaggle public dataset\footnote{https://www.kaggle.com/datasets/kaushiksuresh147/bitcoin-tweets}, and the crypto news information is queried from Nasdaq~\footnote{https://data.nasdaq.com/databases/BICRS/documentation}. Table~\ref{Nasdaq} describes the details of the Nasdaq data.
\begin{table}
\centering
\caption{Data Descriptor of Nasdaq Data}
\scalebox{0.6}{
\begin{tabular}{c c c}
\toprule
\textbf{Column} & \textbf{Description} & \textbf{Type} \\
\midrule
date & The date for which the ratio is calculated (YYYY-MM-DD) & date \\
\midrule
sentiment & Sentiment of the market (e.g., bullish vs. bearish) & text \\
\midrule
indicator & Integer between 0 and 100 representing sentiments from bearish to bullish & integer \\
\bottomrule
\end{tabular}}
\label{Nasdaq}
\end{table}
\subsubsection{Data Preprocessing}
\begin{figure}[htbp]
\centerline{\includegraphics[width=0.7\linewidth]{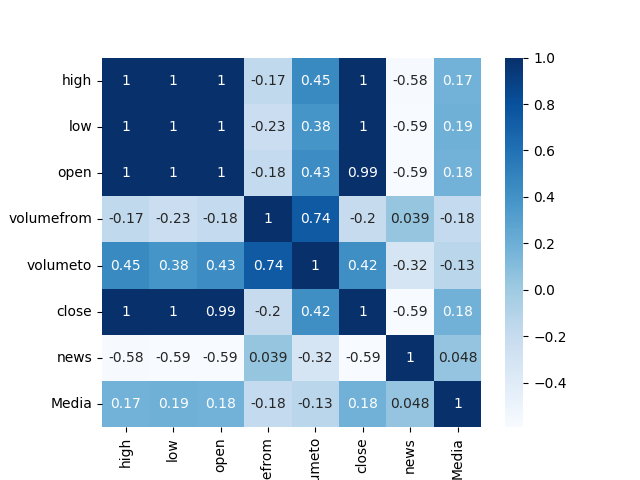}}
\par As shown in Figure~\ref{corr}, the variables \textit{Open, high, low, volume from, news} have large linear correlation with the target variable-\textit{Close}
\caption{Heatmap of feature correlations.}

\label{corr}
\end{figure}

The models we deployed in this paper are multivariate Time Series forecasting models, and the target variable is the daily closing price. We split the last 70-day data as the validation data. To ignore the interruption of irrelevant variables, we first examined the correlation between each variable. As we can observe in Fig.~\ref{corr}, the variables high, low, open price, and volume show a high correlation with the closing price of bitcoin. Considering the high volatility of Cryptocurrency prices, we took the first order percentage difference of the price levels to obtain stationary daily price changes. In our research, we did a comparative study to examine if stationarity enhances the model performances. 

\begin{figure}[htbp]
\centerline{\includegraphics[width=0.7\linewidth]{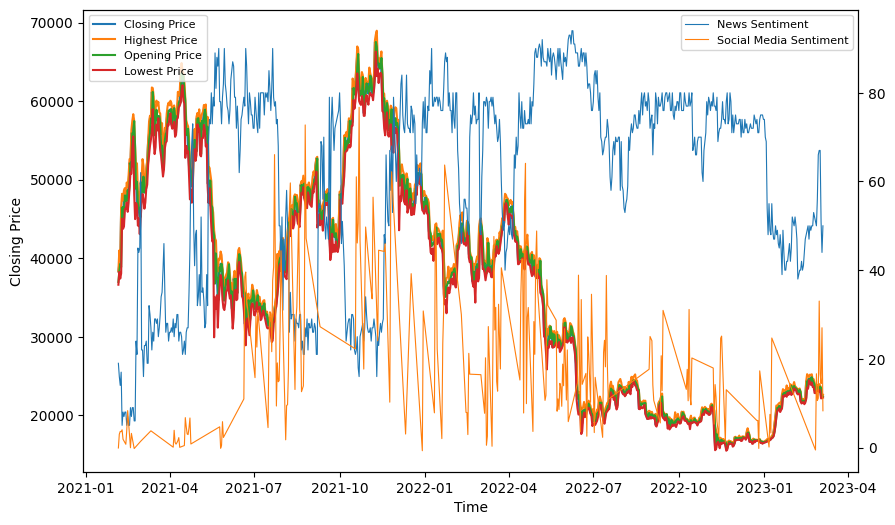}}
\caption{Temporal Data Visualization.}
\label{lineplot}
\par This figure visualizes the raw Bitcoin data in a given period.
\end{figure}

Table~\ref{data} shows the description of each variable of our processed dataset. The variable \textit{Close} is our target variable. In this paper, we introduced the sentiment data produced by CryptoBERT, a pre-trained NLP model to analyze the sentiments of crypto-related social media posts and news as the new modal containing objective and subjective information to improve the performance of our forecasting models~\cite{cryptobert}. The sentiment data was a numerical score within the range of 0 to 1, identifying either bullish or bearish tendencies. Figure~\ref{lineplot} shows the temporal information of our data. After all, we use the basic Min-Max Scaler to rescale all the data with a range from 0 to 1.

\begin{figure*}[t]
\centerline{\includegraphics[width=\textwidth,height=0.23\textheight,keepaspectratio]{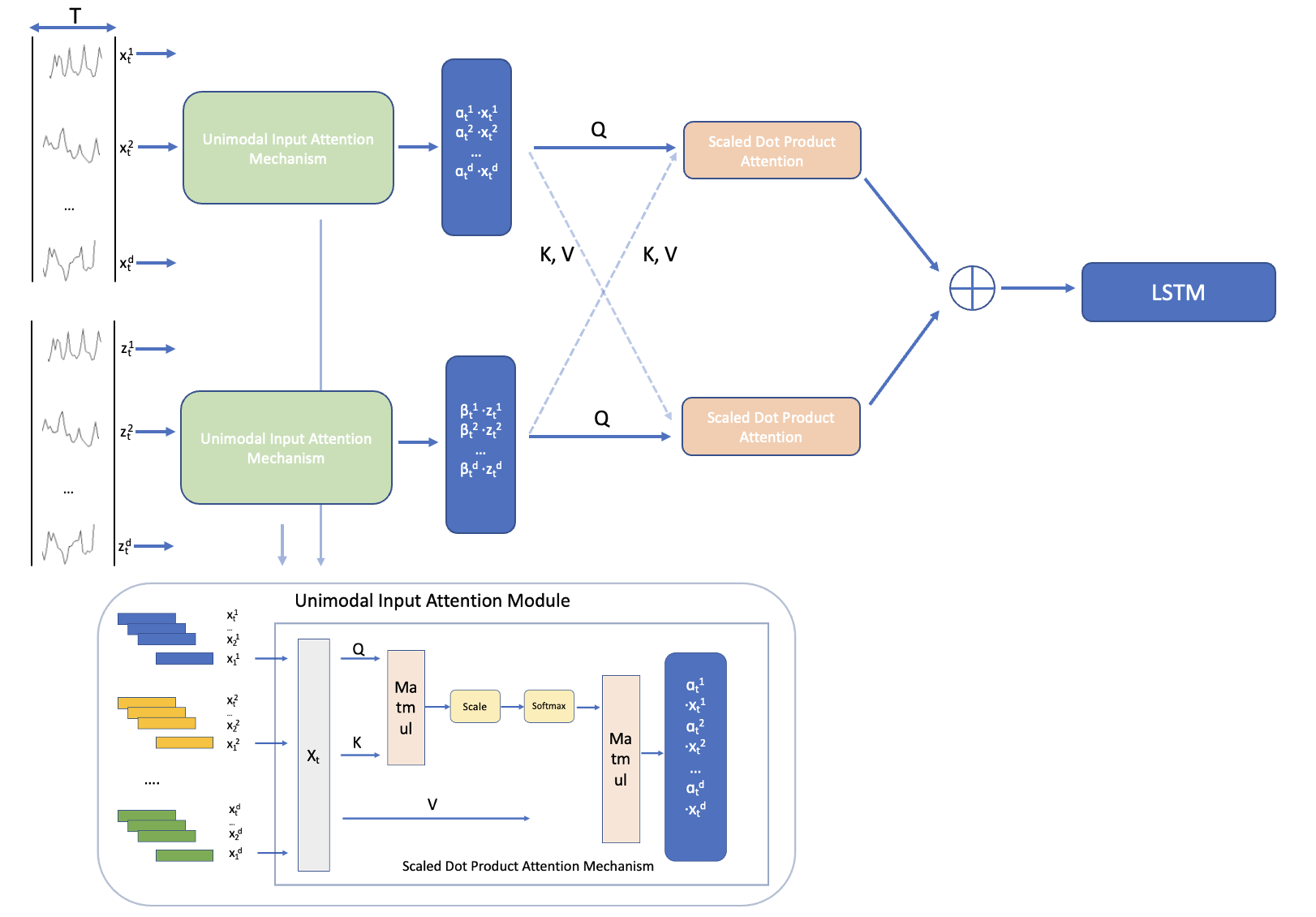}}
\caption{The Dual Attention Mechanism for Multimodal Data Fusion framework.}
\label{model}
\end{figure*}

\begin{figure}[htbp]
\centerline{\includegraphics[width=0.7\linewidth]{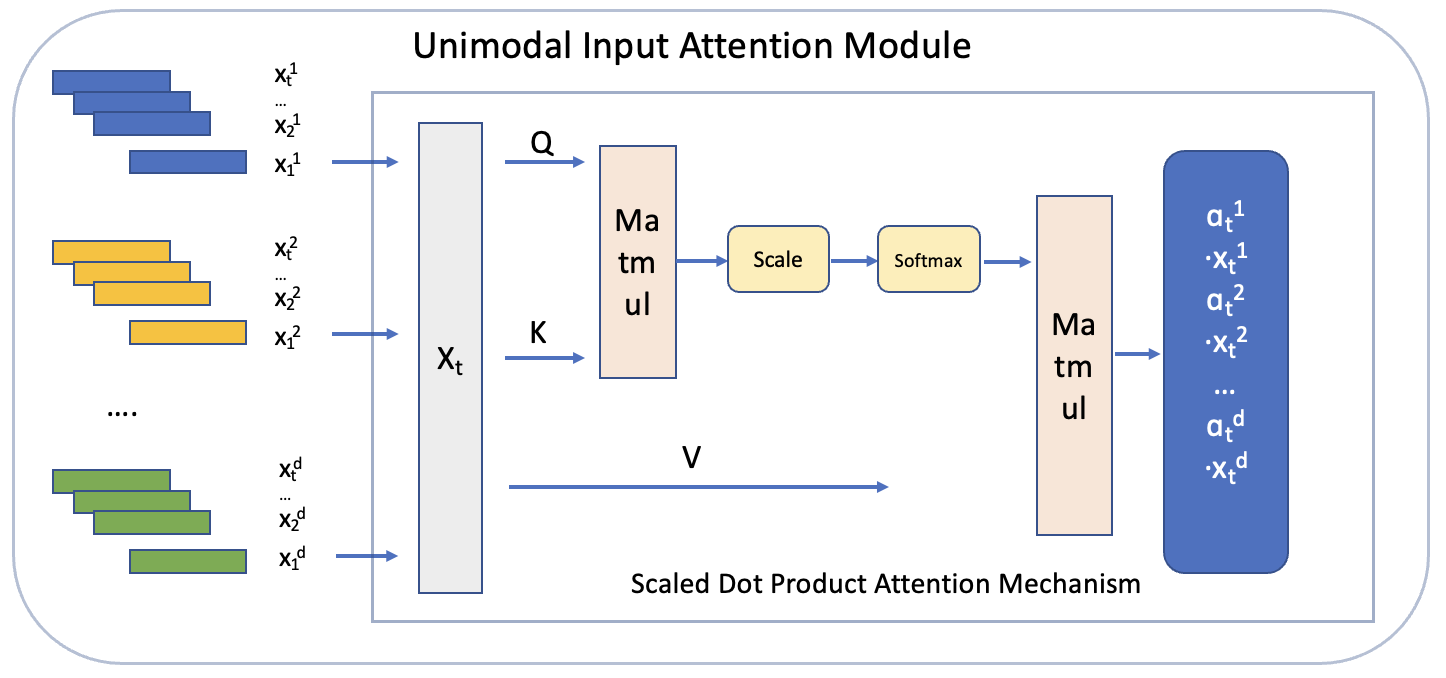}}
\par This figure illustrates the structure of the Unimodal Input Attention Module inside of our DAM model.
\label{Unimodal Input Attention Module}
\caption{Unimodal Input Attention Module.}

\label{Single}
\end{figure}

\begin{table}
      \centering
      \caption{Data Descriptor of Our Processed Dataset}
      \scalebox{0.6}{
      \begin{tabular}{c l c}
        \toprule
        Predictor&{\makecell[c]{Explanation}}&Unit\\
        \midrule
        Opening price& {\makecell[l]{The initial trading price of the cryptocurrency\\ at the beginning of the given day.}}&USD\\
        \midrule
        Closing price& {\makecell[l]{The final trading price of the cryptocurrency\\ at the end of the specified day.}}&USD\\
        \midrule
        {\makecell[c]{Highest/\\Lowest price}} & {\makecell[l]{Highest/Lowest  price at which the Cryptocurrency\\ was traded on the given day.}}&USD \\
        \midrule
        volumefrom & {\makecell[l]{The quantity of shares bought\\ by buyers on the given day.}}&USD \\
        \midrule
        volumeto & {\makecell[l]{The  quantity of shares sold by sellers\\ on the given day.}}&USD \\
        \midrule
        News & {\makecell[l]{The predictor represented the result of the analysis\\ of news contents on the given day.}}&- \\
        \midrule
        Media & {\makecell[l]{The predictor represented the result of the\\ sentiment analysis of social media contents\\ on the given day.\\Positive Media index meant social media critics\\ expected the price to rise.\\Negative Media index meant social media critics\\ expected the price to drop.}}&- \\
        \bottomrule
        
      \end{tabular}}

      \label{data}
    \end{table}   

\subsection{Dual Attention Mechanism for Multimodal Data Fusion}
In this study, we designed an end-to-end Dual Attention Mechanism for Multimodal Temporal Data Fusion. We first introduce the unimodal input attention module, which aims to reconstruct the inputs by deploying the attention weights. Then, we introduce the other attention module to do the cross-modal data fusion. At last, we transfer the new data input to an end-to-end LSTM model to do the final forecasting.

\subsubsection{Unimodal Input Attention Module}
The core idea behind the Unimodal Input Attention Mechanism is to scrutinize the intrinsic dynamics within each modality, effectively capturing intra-modal temporal dependencies.
We have $x^k = (x^k_1, x^k_2, ..., x^k_T)^T \in \mathbb{R}^T \text{ for the first modality}$, 
$z^k = (z^k_1, z^k_2, ..., z^k_T)^T \in \mathbb{R}^T \text{ for the second modality}$, where t denotes the time step and the k denotes the number of temporal series in the certain modal. The input features, either \( x_t^i \) or \( z_t^i \), are transformed into their corresponding queries (Q), keys (K), and values (V) as: $Q = W_q \times I$, $K = W_k \times I$,
$V = W_v \times I$, where $I$ represents $x$ in the first modality and $z$ in the second modality,$W_q$, $W_k$ and $W_v$ are weight matrices for query, key, and value transformations respectively. The attention scores are then computed using:
\begin{align*}
    \text{Attention}(Q, K, V) = \text{Softmax}\left(\frac{Q \times K^T}{\sqrt{d_k}}\right) \times V
\end{align*}
where $d_k$ is the dimensionality of the keys. Then we get the normalized attention weights, which are used to produce a weighted sum of the values, resulting in the output that serves as input for subsequent layers of the model. The detailed structure of the Unimodal Input Attention Module is shown in Figure~\ref{Single}. In the upcoming model training, we could firstly send different input data to this Unimodal module and finally do the multimodal fusion. The cyptocurrency modality data input is a matrix with four dimensions and the sentiment data input is a matrix with two dimensions. 

\subsubsection{LSTM}
Long Short-Term Memory (LSTM) is examined to be a type of recurrent neural network (RNN) architecture that is particularly well-suited for time series forecasting due to its ability to capture and learn patterns in sequential data. Fig.~\ref{LSTM} in appendix~\ref{appendix} shows the architecture of LSTM model. LSTM can remember information over extended time intervals through their cell state, which is controlled by forget, input, and output gates. The mathematical process is in appendx~\ref{appendix}.

\subsubsection{LSTM with Multimodal Fusion}
Considering the cross-modal information correlation, we designed this end-to-end multimodal temporal data fusion mechanism. Given the renewed input data $\hat{x}_T^k$ and $\hat{z}_T^k$, we use another Scaled Dot Product attention mechanism which is similar to our input attention module to do the cross-modal fusion. 

We first compute the query, key, and value representations for each modality:$Q^{fin} = {W^Q_{fin}}\hat{x}_T^k$, $K^{sent} = {W^K_{sent}}\hat{z}_T^k$, $V^{sent} = {W^Q_{sent}}\hat{z}_T^k$, where ${W^Q_{fin}}$, ${W^K_{sent}}$, and ${W^Q_{sent}}$ are the weight matrices for the query, key, and value transformations, respectively. Similar to the previous input attention mechanism, we set the financial modal as Q and the sentiment modal as K, V at the first time. Then, we reverse the order, setting the sentiment modal as Q and concatenate these two outputs. 

With the fused representation $F$, it's primed to be the input to our LSTM model. This LSTM benefits from the comprehensive information encapsulated in $F$ which captures the co-information from both the financial and sentiment perspectives. Given the input $F$ at each time step $t$:$(h_t, s_t) = \text{LSTM}(F_t, h_{t-1}, s_{t-1})$, $\hat{y}_t = W_o h_t + b_o$.

\subsection{Evaluation Metrics}
Our models are evaluated using two metrics: Median Absolute Error (MAE) and Mean Absolute Percentage Error (MAPE). MAE is a universal method to measure forecasting error of timeseries analysis tasks~\cite{hyndman2006another}. MAE calculates the median error across all predictions, providing a measure that is robust to outliers by lessening the influence of anomalously large errors. However, given the high values of cryptocurrency prices, MAE may not explicitly represent the forecast accuracy. Therefore, we utilized MAPE to normalize the errors, providing a percentage-based measure of the average deviation of predictions from actual values, thus offering a clearer perspective on the proportional magnitude of the forecasting errors.

\section{Results}
\label{Results}
In this section, we will introduce our experiment results. We did a comparative study to compare our model with other outstanding models for this task. Also, we explored the explainability of our model using an Ablation Study.
\subsection{Comparative Study}
Table~\ref{result} presents the empirical outcomes derived from the experimentation involving the employed models. Except from our model, the multimodality implemented are basic modal concatenation. We employ the NeuralProphet model as detailed in Triebe et al. (2021)~\cite{triebe2021neuralprophet}, serving as our baseline model. This approach has demonstrated significant enhancements over the well-known Prophet model, particularly in short to medium-term forecasting scenarios. Our Dual Attention Mechanism-LSTM (DAM-LSTM) model outperforms the NeuralProphet model and other selected models and shows a 20 percent improvement on the same LSTM architecture but with the simple concatenation fusion. Interestingly, the technologically more sophisticated Temporal Fusion Transformer (TFT) model, despite its complexity and great performance in long-term forecasting, does not evince commensurate advancements in our dataset. From a global perspective, notably, integrating sentiment data shows a salient impact on predictive performance. The application of multimodal data confers an elevation of prediction by approximately 20 percent for the TFT model and 5 percent for the LSTM model.

Furthermore, the table presents the influence of stationary data input on the forecasting performances of the models. The second column \textit{MAE\textsuperscript{*}} presents the forecasting performances with the stationary data input. As we can observe, the stationary data input improves the LSTM-based model by nearly 50\% percent. However, it does not show much significant improvement on the TFT model.

\begin{table}[htbp]
\centering
\begin{minipage}[t]{0.45\textwidth} 
\centering
\caption{Experiment Results}
\label{result} 
    \scalebox{0.6}{
    \begin{tabular}{c c c c c} 
        \toprule
        Model & \multicolumn{3}{c}{Results} & IsMultimodal \\ 
        \cmidrule(lr){2-4} 
        & MAE & MAE\textsuperscript{*} & MAPE \\ 
        \midrule
        \makecell{\textbf{DAM-LSTM}} & \textbf{719.82} & \textbf{431.86} & \textbf{0.0297} & Yes \\
        \midrule
        \makecell{LSTM} & 837.6 & 491.58 & 0.0377 & Yes \\
        & 863.63 & 501.26 & 0.0392 & No \\
        \midrule
        \makecell{NeuralProphet} & - & 1538.40 & 0.061 & No \\
        \midrule
        \makecell{CNN-LSTM} & 908.86 & 550.73 & 0.0403 & Yes \\
        & 887.65 & 500.15 & 0.0417 & No \\
        \midrule
        \makecell{TFT} & 4523.8740 & 4320.2724 & 0.1936 & Yes \\
        & 6324.7640 & 6158.1635 & 0.2368 & No \\
        \bottomrule
    \end{tabular}}
\end{minipage}
\hfill 
\begin{minipage}[t]{0.45\textwidth}
\centering
\caption{Results of Different Fusion Combinations}
\label{result2} 
    \scalebox{0.6}{
    \begin{tabular}{c c c}
        \toprule
        \multicolumn{3}{c}{Results} \\
        \cmidrule(lr){1-3}
        MAE & & Attention Layer Combination \\
        \midrule
        837.6 & & No any Attention module \\
        790.47 & & No intra-modal attention module \\
        891.25 & & No cross-modal attention module \\
        719.82 & & Dual Attention module \\
        \bottomrule
    \end{tabular}}
\end{minipage}
\end{table}

\subsection{Ablation Study}
Table \ref{result2} provides the results of our Ablation experiments. To examine the performance of each layer of our model, we separately removed each module (each attention module). When we remove the certain attention module, we use the feature concatenation instead. As we can observe from the table, the cross-modal attention layer outweighs the cross-modal attention module. Surprisingly, when we removed the cross-modal attention module, the performance of our model decreased. Consequently, it is discernible that the adoption of our dual attention module engenders a discernible enhancement in the predictive efficacy of our models. We find it reasonable to state that our dual attention module performs better at integrating information from different aspects. 

\section{Discussion}
\label{discussion}
As introduced in the last section, our ablation study finds that when there is no cross-modal attention module, the performance of the model decreases unexpectedly. One possible reason is that as the sequence becomes longer, the LSTM model deteriorates rapidly at a certain point~\cite{cho2014properties}. Researchers proved that using an attention layer before an LSTM model solves this problem efficiently~\cite{qin2017dual}.

While our investigation assessed the effectiveness of multimodality in cryptocurrency market prediction, further exploration is needed to elucidate the underlying rationale behind the performance variations observed across distinct data fusion combinations. Since there exists a time lag between the news report and the market volatility and people's reflections on social media, we examined the correlation between each variable considering the time lags. We can observe from Fig.~\ref{matrice} in the appendix that when the time lag increases (from time lag = 5 to time lag = 30), the correlation between news and close price, news, and social media goes higher. In other words, the influence of news exhibits a hysteresis effect, and this particular characteristic could be taken into account when tackling multimodal cryptocurrency forecasting assignments. However, we found that even though the correlation between news and social media went higher (from 0.058 to 0.11) after involving the hysteresis, our Fisher z-transform tested that there was no significant correlation between news data and social media data (See Table~\ref{tab:correlation_fisher_pvalues}).

\begin{table}[ht]
\centering
\caption{Correlation and Fisher z-transformed p-value for different time lags.}
 \scalebox{0.8}{
\begin{tabular}{|c|c|c|}
\hline
Time Lag & Correlation & Fisher z-transformed p-value \\
\hline
5 & 0.0223 & 0.620 \\
\hline
10 & 0.0225 & 0.624 \\
\hline
15 & 0.0558 & 0.876 \\
\hline
20 & 0.0650 & 0.738 \\
\hline
25 & 0.0870 & 0.448 \\
\hline
30 & 0.128 & 0.120 \\
\hline
32 & 0.124 & 0.142 \\
\hline
35 & 0.111 & 0.222 \\
\hline
40 & 0.088 & 0.432 \\
\hline
\end{tabular}}

\label{tab:correlation_fisher_pvalues}
\end{table}
\vspace{12pt}

Our research corroborates the influence of sentiment on cryptocurrency market fluctuations. Prior studies have demonstrated the positive effect of social media and news sentiments on markets like Blockchain and Ethereum~\cite{ranasinghe2021twitter,youssfi2023bitcoin}. The implications of our findings extend beyond these specific examples, offering applicability in broader financial contexts.

Our study also uncovered a notable finding regarding the underperformance of transformer models, despite their use of attention mechanisms. One popular explanation of the underperforming transformer-based models in this task is the limited data training size. While the feature number is relatively large, the predictive ability deteriorates with the small training data size. That is, in our research, after the data preprocessing, the 700 training data size is not sufficient for forecasting 70 data points. Researchers also investigated several state-of-the-art transformer models designed for Time Series forecasting and concluded that most of the time, the models perform better when the layer number ranges from 3 to 6~\cite{wen2022transformers}. However, the TFT model we use contains a total of 11 layers, including self-attention layers and feed-forward layers. Therefore, the TFT model may not be the most suitable transformer model for this task. 

Furthermore, our research shows the potential to be a universal model that could be applied to different cryptocurrencies. Our DAM model will contribute to the future investigation into the cryptocurrency's volatility issues which is a significant problem practically and ethically in the decentralized science field~\cite{kshetri2022policy,ding2022desci}. Furthermore, our model is more explainable to indicate the effect of public sentiment on the financial market trend than other models that used simple data concatenation. Our improved cryptocurrency forecasting can significantly aid decentralized science (DeSci) by enabling strategic planning and efficient integration of blockchain technologies. This foresight minimizes obsolescence risks and optimizes operational efficiency. Additionally, accurate market predictions help DeSci projects manage funding strategies and financial risks more effectively, ensuring better resource allocation in the dynamic digital asset landscape.

\section{Conclusion and Future Work}
\label{conclusion and future work}
In this study, we have embraced a Dual Attention Mechanism to achieve multimodal fusion, with the primary contributions being the integration of news and social media data to gauge sentiment, the investigation into the effects of including stationary time-series data on model forecasting accuracy, the use of the advanced CryptoBERT for analyzing cryptocurrency sentiment, and the implementation of a comprehensive end-to-end Dual Attention Mechanism for Multimodal Fusion for the final forecasting task. Nonetheless, this research is not without its limitations and there is room for enhancement. We address these limitations and explore avenues for future work in the following section.
\subsection{Generality Examination}
Due to the limited data sources of other cryptocurrencies, especially the sentiment raw data, our research is not tested for a large scale of cryptocurrency data. In the future, manual data mining needs to be done to get the sentiment data and we will test the generality of our DAM model. 
\subsection{Model Selection}
As substantiated by empirical investigations, it has been established that linear models exhibit a comparatively superior efficacy when employed in Time Series tasks as opposed to transformer-based models or other deep learning models. Further exploration could be undertaken by investigating alternative potent models, one such example being the DLinear model proposed by Wen et al. (2022)\cite{wen2022transformers}.

\subsection{Stationary Time-Series}
In this paper, we only introduced a basic method to transform the data to a stationary form. However, researchers postulated more advanced methods that have been tested significantly in deep learning methods. For example, Kuznetsov et al. suggested a convex learning objective in deploying deep-learning models~\cite{kuznetsov2020discrepancy}. Arik et al. put forward a Self-Adaptive Forecasting architecture to enhance deep learning models' performances on non-stationary data~\cite{arik2022self}. In the future study, we might look for more efficient ways to treat the non-stationary data.

\subsection{Model Architecture}
Scholarly investigations have corroborated the presence of inherent seasonality within cryptocurrency dynamics~\cite{kaiser2019seasonality} and researchers postulated that the process of seasonal decomposition constitutes a pivotal facet within the domain of time series forecasting~\cite{wen2020fast}. Drawing from these insights, a potential avenue for enhancing our forthcoming experimental models lies in integrating a dedicated decomposition layer~\cite{bandara2020lstm}. This layer would be designed to effectively capture the intricate interaction between seasonality and underlying trends. 

Also, we noticed that the performance of our dual attention mechanism decreases when the time spanning of our data increases. This situation results in the traditional attention's memory bottleneck and it could be solved by using Flash Attention, which loads keys, queries, and values a single time, integrates the processes of the attention mechanism, and then stores them back~\cite{dao2022flashattention}. However, since it has been widely tested on language models/tasks but not the typical time series data, more experiments need to be done in the future. 

\bibliographystyle{ledgerbib}
\bibliography{tempbib}

\newpage 	

\appendix
\label{appendix}
\setcounter{section}{0}
\section{Data preprocessing}

In this section, we continue to introduce the structure of a basic LSTM model and the mathematical background of the Fisher Transform.

\section{Representative Research}
\begin{table}[!htbp]

      \centering
      \caption{Representative Research on the Multimodal Stock or Cryptocurrency Prediction}
      \scalebox{0.6}{
      \begin{tabular}{c c c c c}
        \toprule
        Category & Reference & Model & Result& IsMultivariate\\
        
        \midrule
        {\makecell[c]{statistical}} & \cite{koki2022exploring} & HMMs & - &- \\
        \cmidrule(l r){2-5}
         & \cite{ariyo2014stock} & ARIMA & S.E 0.7872&No\\
        \cmidrule(l r){2-5}
        & \cite{sardelich2018multimodalstock}&GRACH & MSE $2.09*10^{-5}$& Yes\\
        \cmidrule(l r){2-5}
        & \cite{ibrahim2020bitcoin}&VAR & MAPE $19.88*10^{-2}$& Yes\\
        \midrule
        ML model & \cite{rathan2019crypto} & Decision tree & -&No\\
        \midrule
        {\makecell[c]{DL model}} & \cite{li2020bitcoin} & GAN &RMSE 5.33 &Yes\\
        \cmidrule(l r){2-5}
        &\cite{serafini2020sentiment}&LSTM&MSE 0.00177&Yes\\
        \cmidrule(l r){2-5}
        &\cite{costa2023show}&GNN+Attention&F1 0.801&Yes\\
        \cmidrule(l r){2-5}
        &\cite{phaladisailoed2018machine}&GRU&MSE $2*10^{-5}$&Yes\\
        \cmidrule(l r){2-5}
        &\cite{bhatt2023machine}&LSTM&F1 0.82 & Yes\\
        \cmidrule(l r){2-5}
        &\cite{kim2019forecasting}&CNN+LSTM&MAPE 0.0347&Yes\\
        \cmidrule(l r){2-5}
        &\cite{li2020bitcoin}&CNN+LSTM&MAPE 0.0235&Yes\\
        \cmidrule(l r){2-5}
        &\cite{critien2022bitcoin}&Bi-LSTM&-&Yes\\
        \cmidrule(l r){2-5}
        &\cite{sridhar2021multi}&Transformer&MAE 667&Yes\\
        \cmidrule(l r){2-5}
        &\cite{politis2021ether}&LSTM-TCN&MAPE 0.036&Yes\\
        
        \bottomrule
        \label{review}
      \end{tabular}}
      
     \par Table~\ref{review} provides the representative research of Multimodal stock or Cryptocurrency Prediction tasks. The research papers are categorized based on their model types. Also, this table identifies if one's model is a multivariate forecasting model.
    \end{table}
\section{LSTM Process and Fisher Transform}

\vspace{12pt}

\begin{figure}[!htbp]
\centerline{\includegraphics[width=\linewidth]{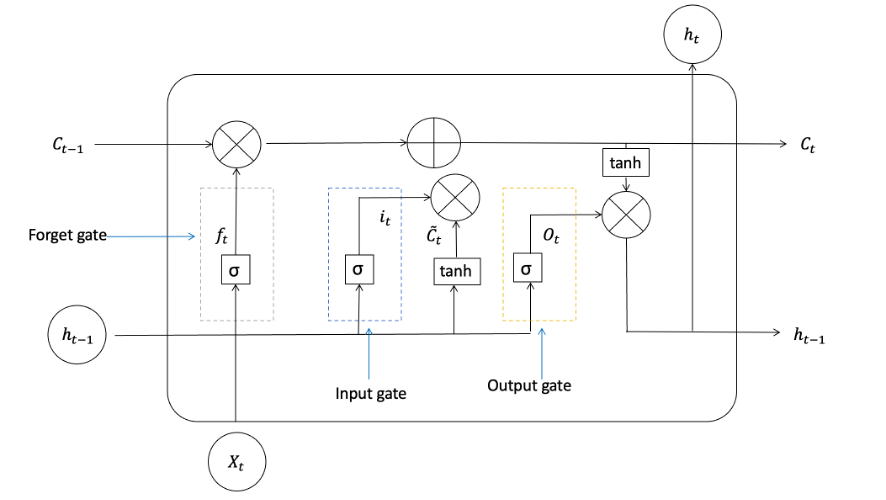}}
\caption{LSTM architecture.}
\label{LSTM}
\end{figure}

Given an input sequence \(x = [x_1, x_2, \ldots, x_T]\), where \(T\) is the length of the sequence, an LSTM cell processes each element \(x_t\) in the sequence to produce an output \(h_t\) at each time step:
\begin{itemize}

    \item Input Gate and Candidate Cell State:
    \begin{align*}
    i_t &= \sigma(W_i \cdot [h_{t-1}, x_t] + b_i) \quad \text{(Input Gate)} \\
    \tilde{C}_t &= \tanh(W_C \cdot [h_{t-1}, x_t] + b_C) 
      \text{(Candidate Cell State)}
   \end{align*}
   
    \item Forget Gate and Previous Cell State:
    \begin{align*}
   f_t &= \sigma(W_f \cdot [h_{t-1}, x_t] + b_f) \quad \text{(Forget Gate)} \\
   C_t &= f_t \cdot C_{t-1} + i_t \cdot \tilde{C}_t \quad \text{(Update Cell State)}
   \end{align*}

   \item Output Gate and Hidden State:
   \begin{align*}
   o_t &= \sigma(W_o \cdot [h_{t-1}, x_t] + b_o) \quad \text{(Output Gate)} \\
   h_t &= o_t \cdot \tanh(C_t) \quad \text{(Hidden State)}
   \end{align*}
    
\end{itemize}

The hidden state \(h_t\) at each time step can be used to make predictions for the next time step or be passed as the context for subsequent LSTM cells in a multi-layer architecture. By processing the input sequence step by step, an LSTM cell can capture and learn intricate temporal patterns in time series data, making forecasting future values based on historical observations practical.

This section introduces the statistical techniques employed to analyze the relationships between variables in our study.

As shown in \ref{corrmath}, the Pearson correlation coefficient measures the extent of a linear association between two continuous variables. It varies between -1 and 1, with 1 indicating a complete positive linear correlation, 0 indicating no linear correlation, and -1 representing a complete negative linear correlation.
\begin{equation}
    r = \frac{\sum_{i=1}^{n} (X_i - \bar{X})(Y_i - \bar{Y})}{\sqrt{\sum_{i=1}^{n} (X_i - \bar{X})^2 \sum_{i=1}^{n} (Y_i - \bar{Y})^2}}
\label{corrmath}
\end{equation}

Where:
\begin{itemize}

\item \quad \(n\) is the number of data points.
\item \quad\(X_i\) and \(Y_i\) are the individual data values.
\item \quad\(\bar{X}\) and \(\bar{Y}\) are the means of \(X\) and \(Y\) respectively.

\end{itemize}

The Fisher $z$-transform is employed to address the limitations of raw correlation coefficients. For example, when the sample correlation coefficient $r$ approaches one or $-1$, its distribution becomes significantly skewed, posing challenges in estimating confidence intervals and conducting significance tests for the population correlation coefficient $\rho$~\cite{fisher1915frequency}. This transformation involves applying the natural logarithm to the ratio of $1$ plus the correlation coefficient ($r$) to $1$ minus the correlation coefficient:
\begin{equation}
    z = \frac{1}{2} \cdot \ln \left( \frac{1+r}{1-r} \right)
\end{equation}
where:
\begin{itemize}
    \item \quad $z$ represents the transformed correlation coefficient (Fisher $z$-score).
    \item \quad $r$ stands for the raw correlation coefficient.
\end{itemize}

\section{Correlation matrics}

\begin{figure}
     \centering
     \begin{minipage}[b]{0.2\textwidth}
         \centering
         \includegraphics[width=\textwidth]{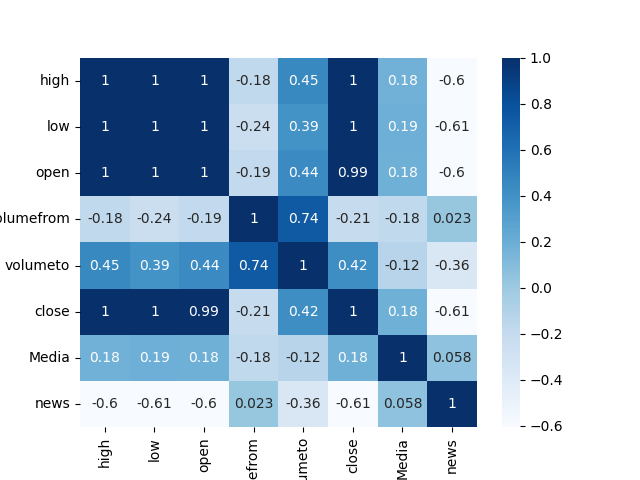}
         \caption{time lag = 5}
         \label{fig:y equals x}
     \end{minipage}
     \hfill
     \begin{minipage}[b]{0.2\textwidth}
         \centering
         \includegraphics[width=\textwidth]{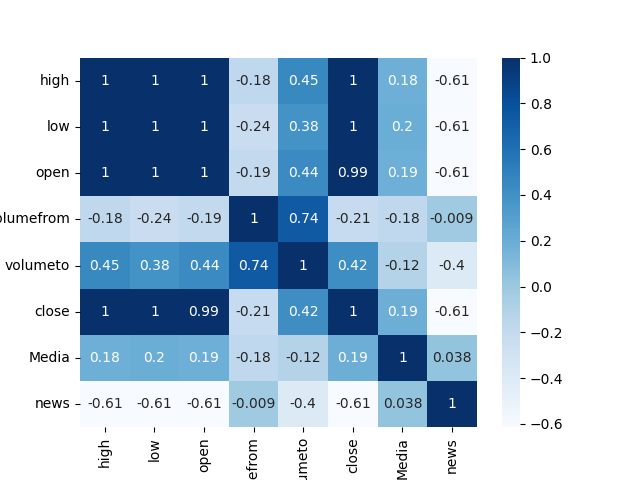}
         \caption{time lag = 10}
         \label{fig:three sin x}
     \end{minipage}
     \hfill
     \begin{minipage}[b]{0.2\textwidth}
         \centering
         \includegraphics[width=\textwidth]{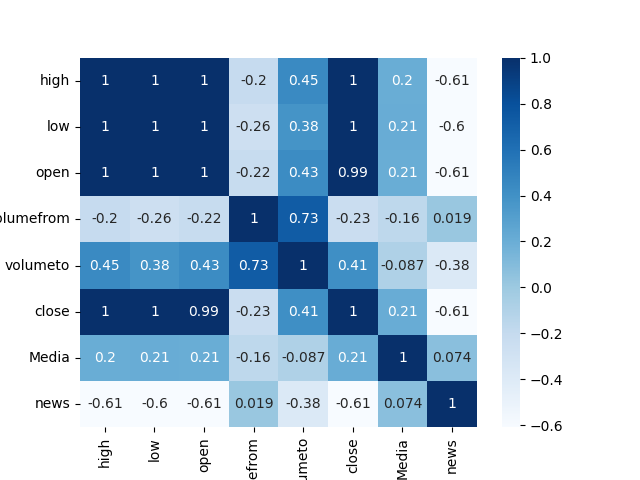}
         \caption{time lag = 20}
         
     \end{minipage}
     \hfill
     \begin{minipage}[b]{0.2\textwidth}
         \centering
         \includegraphics[width=\textwidth]{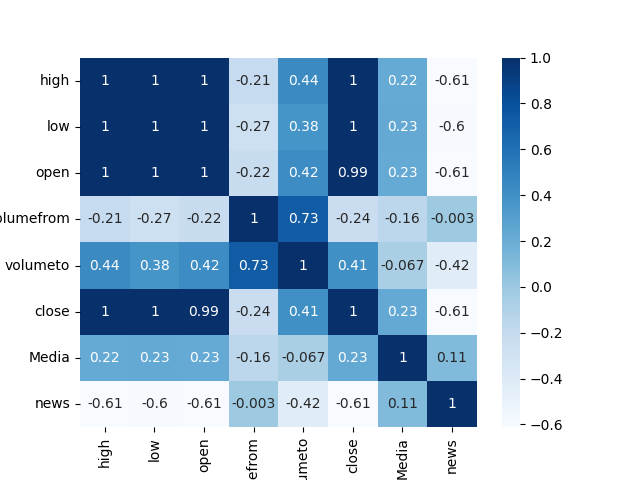}
         \caption{time lag = 30}
         \label{fig:five over x}
     \end{minipage}
     \par These four matrices present the correlation matrix considering the time lags ranging from 5 days to 30 days
        \caption{Correlation matrices considering news hysteresis.}
        
        \label{matrice}
\end{figure}

\end{document}